%
%

\input harvmac

\batchmode
  \font\bbbfont=msbm10
\errorstopmode
\newif\ifamsf\amsftrue
\ifx\bbbfont\nullfont
  \amsffalse
\fi
\ifamsf
\def\IR{\hbox{\bbbfont R}}
\def\IZ{\hbox{\bbbfont Z}}
\def\IF{\hbox{\bbbfont F}}
\def\IP{\hbox{\bbbfont P}}
\else
\def\IR{\relax{\rm I\kern-.18em R}}
\def\IZ{\relax\ifmmode\hbox{Z\kern-.4em Z}\else{Z\kern-.4em Z}\fi}
\def\IF{\relax{\rm I\kern-.18em F}}
\def\IP{\relax{\rm I\kern-.18em P}}
\fi


\def\np#1#2#3{Nucl. Phys. {\bf B#1} (#2) #3}
\def\pl#1#2#3{Phys. Lett. {\bf #1B} (#2) #3}

\def\physrev#1#2#3{Phys. Rev. {\bf D#1} (#2) #3}

\def\rank{{\rm rank}}
\def\Pf{{\rm Pf}}
\def\FI{Fayet-Iliopoulos}

\def\tqi{{\tilde{Q}}^{\tilde i}}

\lref\intseirev{K. Intriligator and N. Seiberg, ``Lectures on
Supersymmetric Gauge Theories and Electric-Magnetic Duality'',
hep-th/9509066, {Nucl. Phys. Proc. Suppl. {\bf 45BC} (1996) 1}.}%

\lref\sem{N. Seiberg, ``Electric-Magnetic Duality in
Supersymmetric Nonabelian Gauge Theories'', \np{435}{1995}{129},
hep-th/9411149.}%

\lref\us{O.~Aharony, A.~Hanany, K.~Intriligator, N.~Seiberg and
M.~J.~Strassler,
``Aspects of $N=2$ Supersymmetric Gauge Theories in Three Dimensions,''
hep-th/9703110.}

\lref\karch{A. Karch, ``Seiberg Duality in Three Dimensions'',
hep-th/9703172.}

\lref\intpou{K. Intriligator and P. Pouliot, ``Exact Superpotentials,
Quantum Vacua and Duality in Supersymmetric $Sp(N_c)$ Gauge
Theories'', \pl{353}{1995}{471}, hep-th/9505006.}

\lref\nati{N. Seiberg, ``Exact Results on the Space of Vacua of Four
Dimensional SUSY Gauge Theories'', 
hep-th/9402044, \physrev{49}{1994}{6857}.}%

\lref\intsei{K. Intriligator and N. Seiberg, ``Mirror Symmetry in Three
Dimensional Gauge Theories,'' hep-th/9607207, \pl{387}{1996}{513}.}

\lref\bhoy{J. de Boer, K. Hori, Y. Oz and Z. Yin, ``Branes and Mirror
Symmetry in $N=2$ Supersymmetric Gauge Theories in Three Dimensions'',
hep-th/9702154.}%

\lref\berkeley{J. de Boer, K. Hori, H. Ooguri and Y. Oz, ``Mirror Symmetry
in Three Dimensional Gauge Theories, Quivers and D-branes'',
hep-th/9611063 \semi
J. de Boer, K. Hori, H. Ooguri, Y. Oz and Z. Yin, ``Mirror
Symmetry in Three Dimensional Gauge Theories, $SL(2,\IZ)$ and D-brane
Moduli Spaces'', hep-th/9612131 \semi
J. de Boer, K. Hori, Y. Oz and Z. Yin, ``Branes and Mirror
Symmetry in $N=2$ Supersymmetric Gauge Theories in Three Dimensions'',
hep-th/9702154.}

\lref\me{O. Aharony, ``Remarks on Non-Abelian Duality in $N=1$
Supersymmetric Gauge Theories'', \pl{351}{1995}{220}, hep-th/9502013.}

\lref\aps{P. C. Argyres, M. R. Plesser and N. Seiberg, ``The Moduli
Space of Vacua of $N=2$ SUSY QCD and Duality in $N=1$ SUSY QCD'',
\np{471}{1996}{159}, hep-th/9603042.}

\lref\egk{S. Elitzur, A. Giveon and D. Kutasov, ``Branes and N=1
Duality in String Theory'', hep-th/9702014.}%

\lref\oogvaf{H. Ooguri and C. Vafa, ``Geometry of $N=1$ Dualities in
Four Dimensions'', hep-th/9702180.}

\lref\leeyi{K. Lee and P. Yi, ``Monopoles and Instantons in Partially
Compactified D-branes'', hep-th/9702107.}%

\lref\hw{A. Hanany and E. Witten, ``Type IIB Superstrings, BPS
Monopoles, and Three Dimensional Gauge Dynamics'', hep-th/9611230.}%


\Title{RU-97-18, hep-th/9703215}
{\vbox{\centerline{IR Duality in $d=3$ $N=2$ Supersymmetric}
        \vskip4pt\centerline{$USp(2N_c)$ and $U(N_c)$ Gauge Theories}}}
\centerline{Ofer Aharony
\foot{oferah@physics.rutgers.edu 
}}
\bigskip\centerline{Department of Physics and Astronomy}
\centerline{Rutgers University}
\centerline{Piscataway, NJ 08855-0849}

\vskip 1in
We suggest IR-dual descriptions for $d=3$ $N=2$ supersymmetric gauge
theories with gauge groups $USp(2N_c)$ and $U(N_c)$ and matter in the
fundamental representation. We relate this duality to the IR duality
of $d=4$ $N=1$ SQCD theories, and in one case also to mirror symmetry.

\Date{3/97}

\newsec{Introduction}

In the past few years, many exact (non-perturbative) results have been
found regarding the moduli spaces and superpotentials of $d=4$ $N=1$
supersymmetric gauge theories (see \intseirev\ for a review). One of
the most interesting results is the existence of pairs of gauge
theories which flow to the same theory in the IR limit, first
discovered in \sem. If one of the pair of theories is IR-free, it
provides a good description of the theory in the IR, while otherwise
they both flow to the same non-trivial fixed point.

Similar dualities, which were called ``mirror symmetries'', were
recently discovered also in $d=3$ supersymmetric theories, both with
$N=4$ supersymmetry and with $N=2$ supersymmetry
\refs{\intsei,\berkeley,\us}. These dualities also relate the IR
behaviors of two different gauge theories, and interchange their Higgs
and Coulomb branches (which are distinguishable in $N=4$ theories but
not in $N=2$ theories). In three dimensions all gauge theories are
strongly coupled in the IR, so dualities between gauge theories never
give effective IR descriptions, but only theories flowing to the same
strongly coupled fixed points.

The moduli spaces of $d=3$ $N=2$ SQCD theories (analyzed for $SU(N_c)$
and $U(N_c)$ gauge theories in \us, and for $USp(2N_c)$ gauge theories
in \karch), show many similarities to their $d=4$ $N=1$ counterparts
(except for the existence of an additional Coulomb branch). As in the
$d=4$ $N=1$ theories (up to a shift of $N_f$ by one), for small number
of flavors there is no supersymmetric vacuum, for $N_f=N_c-1$ (in
$SU(N_c)$ and $U(N_c)$ theories, or $N_f=N_c$ in $USp(2N_c)$ theories)
there is a smooth moduli space, for $N_f=N_c$ ($N_f=N_c+1$) there is a
dual description of the origin of moduli space using only chiral
multiplets, while for larger values of $N_f$ there is a moduli space
with a singularity at the origin, for which no dual description is
known. The $d=4$ $N=1$ results suggest that a dual description should
exist for these singularities at the origin of moduli space, and we
will show that (at least for $U(N_c)$ and $USp(2N_c)$ gauge theories)
this is indeed the case. A similar (but not identical) duality for
$USp(2N_c)$ gauge theories was recently suggested in \karch.

We should emphasize that the dual descriptions that we describe here
are not expected to be unique (thus, the word ``duality'' is perhaps a
misnomer). For instance, all the theories we discuss here have also a
``mirror'' description (which one can flow to from the $N=4$
``mirror'' description) which is not identical with their ``dual''
description (except in special cases). The full set of relationships
between $d=3$ $N=2$ theories has yet to be explored. The ``mirror''
symmetry for Abelian theories was interpreted as exchanging bound
states of the original ``electric'' variables with Nielsen-Olesen
vortices in \us. Perhaps similar interpretations exist also for other
dualities, which might shed light also on the duality in $d=4$ $N=1$
theories.

The $d=4$ $N=1$ dualities were given an interpretation in terms of
brane constructions (generalizing the brane constructions of \hw) in
\refs{\egk,\oogvaf}. These constructions may easily be generalized to
the $d=3$ $N=2$ case, and they suggest dualities which are similar to
the ones described here (they involve the same gauge groups and
charged matter content, but it is difficult to see the superpotentials
involving the quantum variables on the Coulomb branch in the brane
construction). The brane construction also suggests a duality between
$SO(N_c)$ gauge groups and $SO(N_f-N_c+2)$ gauge groups, which we will
not discuss here. As discussed in \oogvaf, the brane construction
cannot be viewed as a proof of these dualities, since it only shows
that there is a smooth interpolation between the two theories (which
is not always an IR-equivalence, even if both theories are
asymptotically free). It would be interesting to understand the
conditions for a brane construction of this type to give a consistent
field theory duality.

\newsec{Duality in $USp(2N_c)$ gauge theories}

In this section we propose an IR-dual description for $d=3$ $N=2$
$USp(2N_c)$ gauge theories with $2N_f$ fundamental flavors. The
duality is similar in spirit to the Seiberg duality in four dimensions
\refs{\sem,\intpou}, and is connected to it when we discuss the $d=4$
theory on a circle of finite radius, as described below. Another
duality for these theories was proposed in \karch, which is similar to
ours but involves an extra $SU(2)$ symmetry whose role is not clear
(and which seems to destroy the duality when included in the
dynamics). Generally, we could have many different theories flowing to
the same IRFP, so many other dual descriptions might also exist.

The ``original'' theory will be a $USp(2N_c)$ $N=2$ gauge theory, with
chiral multiplets $Q_i$ ($i=1,\cdots,2N_f$) in the fundamental ($\bf
2N_c$) representation. The gauge invariant operators parametrizing the
Higgs branch are $M_{ij} = Q_i
Q_j$, and the Coulomb branch (after part of it is lifted by
instantons) is parametrized by a chiral superfield
$Y$ (as in \refs{\us,\karch}). The quantum-corrected global charges of
the various fields are (with a convenient choice for the $U(1)_R$
symmetry) :
\eqn\origglob{\matrix{
\quad & U(1)_R & U(1)_A & SU(2N_f) \cr
Q & 0 & 1 & {\bf 2N_f} \cr
M & 0 & 2 & {\bf N_f(2N_f-1)} \cr
Y & 2(N_f-N_c) & -2N_f & {\bf 1}. \cr}}

The ``dual'' theory we propose is a $USp(2(N_f-N_c-1))$ gauge theory,
with $2N_f$ chiral multiplets $q_i$ in the fundamental representation,
and with additional singlet chiral multiplets $M$ and $Y$
(corresponding to the fields $M$ and $Y$ defined above). The Coulomb
branch parameter of the ``dual'' theory will be denoted by ${\tilde
Y}$, and we will choose the quantum numbers of the dual fields such
that a $M q q$ superpotential is possible. This leads to the following
charge assignments for the new fields appearing in the ``dual''
theory :
\eqn\dualglob{\matrix{
\quad & U(1)_R & U(1)_A & SU(2N_f) \cr
q & 1 & -1 & {\bf 2N_f} \cr
{\tilde Y} & -2(N_f-N_c-1) & 2N_f & {\bf 1}. \cr}}
We suggest that the ``dual'' theory has a superpotential of the
form\foot{All scales are set to one in this paper for simplicity.}
\eqn\spot{W = M_{ij} q_i q_j + Y \tilde{Y},}
which is consistent with the global symmetries (note that in
non-Abelian theories there is no exact $U(1)_J$ symmetry acting on the
``dual photons'' $Y$). As in $d=4$ $N=1$ duality, performing the
duality transformation twice returns us to the original theory.
It is not clear how to define the
superpotential \spot\ microscopically, since $\tilde{Y}$ is not a
gauge-invariant variable (but only an effective variable on the
Coulomb branch). Perhaps the chiral superfield $S = W_{\alpha}^2$ of
the ``dual'' theory, which is related to $\tilde{Y}$ (they depend on the
same ``fundamental'' fields), can be used for such a microscopic
definition of the ``dual'' theory. In any case, since we will only be
discussing the IR behavior of the theory here, we will be content with
using the superpotential \spot.

As in $d=4$ $N=1$ theories, we can test this duality in various
ways. The global symmetries and gauge invariant chiral superfields are
obviously the same in the two theories. Next, let us compare the
moduli spaces of the two theories. In the ``original'' theory, for
$Y=0$ the meson $M$ can obtain a VEV with $\rank(M) \leq 2N_c$, while
for $Y \neq 0$ it satisfies $\rank(M) \leq 2(N_c-1)$. In the ``dual''
theory, naively $M$ can obtain any VEV (up to rank $2N_f$). The
equation of motion of $\tilde Y$ appears to set $Y=0$, but we should
remember that for $N_f > N_c+1$ we have no good description of the
region of moduli space near the origin ($\tilde{Y}=0$) in which
$\tilde{Y}$ is a fundamental variable, so we cannot simply use this
equation of motion. Using the ``effective'' superpotential on the
moduli space, of the form $W \sim (\tilde{Y}
\Pf(q_iq_j))^{1/(N_f-N_c-1)}$
(which gives a good description of the region of moduli space away
from the origin), suggests that $Y$ can generically obtain any VEV in
the ``dual'' theory (as long as $\rank(M) < 2N_c$, as discussed
below). However, since $Y$ is a singlet chiral superfield in the
``dual'' theory, we can use its equation of motion and find
$\tilde{Y}=0$, so the dual Coulomb branch is lifted.  Now, if we give
$M$ a VEV of rank $2N_c$, we are left (using the superpotential) with
$2(N_f-N_c)$ massless quarks, and the low-energy theory then has a
dual description in terms of a superpotential \karch\ of the
form $W = -{\tilde Y} \Pf(q_iq_j)$. Together with the original
superpotential
\spot\ 
this now sets $Y = \Pf(q_iq_j) = 0$, in agreement with the
``original'' theory.
For $\rank(M) = 2(N_c+1)$, instantons in the low energy theory
generate a constraint of the form ${\tilde Y} \Pf(q_iq_j) = 1$
\karch, which is inconsistent with the superpotential, so
there are no supersymmetric vacua, as in the ``original''
theory. Thus, the moduli spaces of the two theories are the same.

It is easy to check that, just as in four dimensions, the duality
behaves in an appropriate way under complex mass perturbations
(reducing $N_f$) and
flows along its Higgs branch (reducing $N_f$ and $N_c$). 
The duality obeys the parity anomaly
matching conditions discussed in \us. The effect of adding real
masses, which can also be identified in the ``dual'' theory since they
can be thought of as background global vector fields \us, is more
complicated and will not be discussed here. 

The ``dual'' theory exists for $N_f > N_c+1$. For $N_f=N_c+1$, a dual
description without gauge fields exists, as discussed in \karch. If
we start with the $N_f=N_c+2$ theory, whose dual is a $USp(2)$ theory,
and add a mass term $mM$ for one of the quarks, the dual gauge
theory breaks completely, and after integrating out the (now massive)
dual quarks and $\tilde{Y}$, the global symmetries guarantee that we
will go over to the known dual description using just $Y$ and $M$
\karch, which is $W = -Y \Pf(M)$.

Our $d=3$ results can be connected with
the $d=4$ duality for the same gauge groups
\intpou. As discussed in \refs{\us,\karch}, the description of the
$d=4$ $N=1$ theory compactified on a circle of radius $R$ is the same
as that of the $d=3$ theory, except for an additional ``twisted
instanton'' \leeyi\ contribution to the superpotential of the form $W
= \eta Y$, where $\eta \sim e^{-1/Rg^2} \sim e^{-1/g_4^2}$. Due to the
superpotential of the ``dual'' theory, we do not know how to analyze
this theory directly at finite radius. However, if we add the
correction to the superpotential of the ``original'' theory also to
the ``dual'' theory, we find that for finite radius $\tilde{Y} =
-\eta$, so the ``dual'' theory is always on its Coulomb branch, where
classically the gauge symmetry is broken to $USp(2(N_f-N_c-2))\times
U(1)$. In the $d=4$ limit, we can integrate out the fields $Y$ and
$\tilde{Y}$, and we are left with the duality of $d=4$ \intpou : a
dual $USp(2(N_f-N_c-2))$ gauge group with a superpotential $W = M_{ij}
q_i q_j$ (the remaining $U(1)$ gauge field is free and decouples in
this limit).

\newsec{Duality in $U(N_c)$ gauge theories}

The duality for $U(N_c)$ gauge theories is similar to the one
described in the previous section. The ``original'' theory has $N_f$
chiral multiplets $Q_i$ in the $\bf N_c$ representation, and $N_f$
chiral multiplets $\tqi$ in the $\bf{\overline{N_c}}$
representation. The Higgs branch may be parametrized by the
gauge-invariant superfields $M_i^{\tilde i} = Q_i \tqi$. The Coulomb
branch remaining after the instanton corrections is now parametrized
by two chiral superfields, $V_+$ and $V_-$ \us, and the
quantum-corrected global charges are :
\eqn\origglob{\matrix{
\quad & U(1)_R & U(1)_A & SU(N_f) & SU(N_f) & U(1)_J \cr
Q & 0 & 1 & {\bf N_f} & {\bf 1} & 0\cr
{\tilde Q} & 0 & 1 & {\bf 1} & {\bf \overline{N_f}} & 0\cr
M & 0 & 2 & {\bf N_f} & {\bf \overline{N_f}} & 0\cr
V_{\pm} & N_f-N_c+1 & -N_f & {\bf 1} & {\bf 1} & \pm 1. \cr}}

The quantum moduli space of these theories was discussed in
\us. For $N_f=N_c$, there is a dual description in terms of a
superpotential $W = -V_+ V_- \det(M)$, from which we can flow to
theories with smaller values of $N_f$. For higher $N_f$, the only
superpotential consistent with the global symmetries is $W \sim (V_+ V_-
\det(M))^{1/(N_f-N_c+1)}$, which describes the moduli space correctly
but is singular at the origin. In particular, for $V_+=V_-=0$, $M$
obeys $\rank(M) \leq N_c$, while if one of them is non-zero we find
$\rank(M)
\leq (N_c-1)$, and if both $V_+$ and $V_-$ are non-zero, $\rank(M) \leq
(N_c-2)$.

The ``dual'' theory we propose is a $U(N_f-N_c)$ theory with $N_f$
flavors $q^i$ (in the $\bf{N_f-N_c}$ representation) and ${\tilde
q}_{\tilde i}$ (in the $\bf\overline{N_f-N_c}$ representation), much
as in $d=4$ \sem. As in the previous section, we will take $M$, $V_+$
and $V_-$ to be singlet fields in the ``dual'' theory, and choose the
global charges of $q$ and $\tilde{q}$ to be consistent with a $M q
\tilde{q}$ superpotential. This leads to the quantum-corrected global
charges :
\eqn\dualglob{\matrix{
\quad & U(1)_R & U(1)_A & SU(N_f) & SU(N_f) & U(1)_J \cr
q & 1 & -1 & {\bf \overline{N_f}} & {\bf 1} & 0\cr
{\tilde q} & 1 & -1 & {\bf 1} & {\bf N_f} & 0\cr
{\tilde V}_{\pm} & N_c-N_f+1 & N_f & {\bf 1} & {\bf 1} & \pm 1. \cr}}
We suggest that the superpotential of the ``dual'' theory is
\eqn\dualsupot{W = M_i^{\tilde i} q^i {\tilde q}_{\tilde i} + V_+
\tilde{V}_- + V_- \tilde{V}_+,}
which is consistent with the global symmetries described above. As
in $d=4$ $N=1$ duality, performing the duality transformation twice
returns us to the original theory.

The tests we can perform of this duality are the same as the ones
described in the previous section. The
only small difference is in comparing the Higgs branches of the two
theories. Now, when we give $M$ a VEV of rank $N_c$, we flow at low
energies to the $U(N_f-N_c)$ theory with $N_f-N_c$ flavors, which has
a dual description in terms of a superpotential $W = {\tilde V}_+
\tilde{V}_- \det(q \tilde{q})$. Adding this to \dualsupot, we find
that the equations of motion set $V_+ = V_- = 0$, as in the
``original'' theory. If $\rank(M)=(N_c-1)$ things are a bit more
complicated, since we have no consistent dual description of the low
energy theory in this case. However, using the ``effective'' low
energy superpotential $W \sim (\tilde{V}_+ \tilde{V}_-
\det(q\tilde{q}))^{1/2}$ for this case, we find (using \dualsupot)
that $V_+$ and $V_-$ can be non-zero but $V_+ V_- = 0$, again as
in the ``original'' theory. For $\rank(M) > N_c$ we find, as
before, no stable vacua because of instanton effects. The standard
Coulomb and Higgs branches of the ``dual'' theory are 
obviously lifted by \dualsupot.
Thus, the moduli spaces of the two theories are the same.

An additional deformation we can perform in the $U(N_c)$ case is to
add a \FI\ term to the ``original'' theory. As described in \us, this
term can be thought of as a background $U(1)_J$ vector field, so in
the ``dual'' theory it becomes both a \FI\ term and a real mass term
for $V_+$ and $V_-$. The \FI\ term forces the gauge symmetries of both
theories to be completely broken. The simplest flat direction now
involves taking only $Q$'s to be non-zero in the ``original'' theory
(assuming the \FI\ term is positive) -- this is the ``baryonic'' flat
direction, discussed in \refs{\me,\aps} (and used in \egk\ to relate
$d=4$ $N=1$ dual theories). In the ``original'' theory, the gauge
symmetry is completely broken, some of the $Q$'s are swallowed by the
Higgs mechanism, and $N_c(N_f-N_c)$ $Q$'s and $N_c N_f$ $\tilde Q$'s
remain massless (and free in the IR). In the ``dual'' theory, again
only $q$'s obtain VEVs, breaking the gauge group completely. Some of
the $q$'s are swallowed by the Higgs mechanism, and the $\tilde{q}$'s
and some of the $M$'s are given a mass by the superpotential. We
remain with $N_c (N_f-N_c)$ massless $q$'s and $N_f N_c$ massless
$M$'s, which are again free in the IR, and which may be identified
with the remaining fields of the ``original'' theory \me. The Coulomb
branches of both theories are lifted (including the branch with non-zero
$V_\pm$ in the ``dual'' theory, due to their real masses), so the
duality becomes a trivial IR identification of free fields in this
case (like in $d=4$ \me).

The ``dual'' gauge theory description exists for $N_f > N_c$, while
for $N_f=N_c$ a dual description is known involving just $V_+$,$V_-$
and $M$ with no gauge fields \us. If we start with the $N_f=N_c+1$
theory, whose dual is a $U(1)$ gauge theory, and add a mass term $mM$
for one of the quarks, the ``dual'' gauge theory breaks
completely. Integrating out the massive quarks and $\tilde{V}_{\pm}$
fields, the global symmetries guarantee that we go over to the known
dual description $W = -V_+ V_- \det(M)$ for the $N_f=N_c$ theory.

Flowing to $d=4$ in this case involves adding a ``twisted instanton''
superpotential of the form $W = \eta V_+ V_-$ (for $N_c > 1$). Again,
it is not clear how this term arises in the ``dual'' theory, but we
will assume that it arises there as well as in the ``original''
theory. Now, in the $d=4$ limit, we integrate out the Coulomb branch
fields, and remain with the standard duality of $d=4$ $U(N_c)$ gauge
theories
\sem\ (which follows from the duality of $SU(N_c)$ theories by gauging
the global $U(1)_B$ symmetry). Note that this works even for $N_f =
N_c+1$. In this case we end up in $d=4$ with the ``effective''
descriptions of the $N_f=N_c+1$ SQCD theories
\nati, again with the $U(1)_B$ symmetry gauged (and free in the IR).

For the $U(1)$ gauge theory with $N_f=2$, the ``dual'' theory is also
a $U(1)$ theory with two flavors, and the same is true also for the
``mirror'' description of this theory
\refs{\intsei,\berkeley,\us}. The ``mirror'' description involves a
superpotential of the form $W = S_1 q^1 \tilde{q}_1 + S_2 q^2
\tilde{q}_2$ with two singlets $S_1$ and $S_2$, which are identified
with two of the meson fields of the ``original'' theory (say, $M^1_1$
and $M^2_2$). The ``dual'' description (defined above) differs from
this by having four additional singlet fields in the ``dual'' theory,
$M^1_2$, $M^2_1$, $V_+$ and $V_-$, and by having a manifest
$SU(2)\times SU(2)$ global symmetry (which is not manifest in the
``mirror'' theory). The equations of motion, however, relate these
additional fields to the other ``dual'' fields (for instance, they set
$V_+ V_- \sim \det(q^i
\tilde{q}_{\tilde i})$), so we can think of them as auxiliary fields,
and then the ``dual'' description seems to be very similar to the
``mirror'' description. This relationship suggests that perhaps the
duality we describe here can also be understood in terms of classical
vortex solutions, like the ``mirror'' symmetry \us.

Presumably, it should be possible to generalize this duality also to
$SU(N_c)$ gauge groups, but it is not obvious how to do this. A
duality for $SU(N_c)$ gauge groups may easily be turned into a
$U(N_c)$ duality by gauging a global symmetry, but ``ungauging''
symmetries is more difficult, and the $U(1)$ gauge field has important
dynamics (such as confinement) in $d=3$. Generalizations to other
cases (with or without known $d=4$ dualities) should presumably also
be possible.

\bigskip
\centerline{{\bf Acknowledgments}}

I would like to thank A. Hanany, K. Intriligator and N. Seiberg for
useful discussions. This work was supported in part by DOE grant
DE-FG02-96ER40559.

\listrefs

\end